\documentclass[aps,preprint,showkeys,superscriptaddress]{revtex4-1}
\usepackage{graphicx}
\usepackage{amsmath}
\usepackage{amssymb}
\usepackage{longtable}
 \usepackage{dcolumn}
\usepackage{color}
 \usepackage{natbib}

\begin{document}

\author{D. Asthagiri}
\email{Dilip.Asthagiri@rice.edu}
\affiliation{Department of Chemical and Biomolecular Engineering, Rice University, Houston, TX}
\date{\today}

\title{Thermodynamics of the collapse transition of the all-backbone peptide Gly$_{15}$}



\date\today
\begin{abstract}
Simulations show Gly$_{15}$, a polypeptide lacking any side-chains, can collapse in water.  We assess the hydration thermodynamics in this collapse
by calculating the hydration free energy at each of the end points of the reaction coordinate, here the end-to-end distance ($r$) in the chain.  
To examine the role of the various conformations for a given $r$, we study the conditional distribution, $P(R_g | r)$, of the radius of gyration for
a given value of $r$. $P(R_g|r)$ is found to vary more gently compared to the corresponding variation in the excess hydration free energy. Using this
insight within a multistate generalization of the potential distribution theorem, we calculate a reasonable upper bound for the hydration free energy of the peptide
for a given $r$. On this basis we find that peptide hydration greatly favors the expanded state of the chain, despite primitive hydrophobic 
effects favoring chain collapse. The net free energy of collapse is seen to be a delicate balance between opposing intra-peptide and hydration effects, 
with intra-peptide contributions favoring collapse by a small margin. The favorable intra-peptide interactions are primarily electrostatic in origin, and
found to arise primarily from interaction between C=O dipoles, hydrogen bonding interaction between C=O and N-H groups, and favorable 
interaction between N-H dipoles. 
\end{abstract}

\maketitle

The concept of hydrophobic hydration, the tendency of apolar solutes to disfavor the aqueous phase, informs nearly all aspects of biomolecular self-assembly and  
is commonly accepted as providing the driving force for proteins to fold \cite{Kauzmann:59, chandler:nature05, dill:1990ww,Dill2012}.
However, this rationalization cannot explain recent experimental \cite{Teufel:2011jmb} and simulational \cite{Tran:2008bk, Hu:2010b} observations that oligoglycine, only mildly 
hydrophobic by some accounts \cite{delisi:jmb87,Wilce1995}, also collapses into a non-specific structure in liquid water.  

Experimental studies on the collapse of (Gly)$_n$ and the closely related (GlySer)$_{n}$ polypeptides have attributed the collapse to the formation of intramolecular hydrogen bonds \cite{Kiefhaber:pnas06,Teufel:2011jmb}.  However, an earlier simulation study has suggested that collapse is unlikely to be driven solely by intramolecular hydrogen bonding \cite{Tran:2008bk}. They have instead postulated that the unfavorable cost of creating a cavity to accommodate the  peptide drives the collapse, a picture that is synonymous with hydrophobicity driven collapse. More recent work has implicated the charge ordering and the favorable correlation between the CO groups of the peptide as an important determinant in oligoglycine collapse \cite{Karandur2014,Karandur2015a}. A rigorous analysis of hydration effects in folding of Gly$_{15}$ has not yet been presented. 

Here we explore the hydration thermodynamics of Gly$_{15}$ collapse using the recently developed regularization approach to free energy calculations \cite{weber:jcp11, Weber:jctc12}. This approach makes possible the facile calculation of free energies of hydration of polypetides and proteins in all-atom simulations. Importantly, this approach provides direct quantification of the hydrophilic and hydrophobic contributions to hydration \cite{tomar:jpcb14,tomar:jpcb16}. We complement these studies with evaluation of the excess enthalpy and entropy of hydration as well \cite{tomar:jpcb14,tomar:jpcb16}. Our results show that in contrast to the usual paradigm of water aiding folding by decreasing the mutual solubility of the peptide units comprising the polypeptide chain, hydration in fact drives unfolding in this peptide;  importantly, intra-peptide van~der~Waals and electrostatic interactions are critical in driving Gly$_{15}$ to collapse. Some of the favorable electrostatic interactions are clearly attributable to the formation of hydrogen bonds, as was suspected in the experimental studies \cite{Kiefhaber:pnas06,Teufel:2011jmb}. 

\section{Methods}

Gly$_{15}$ was constructed with capped ends and solvated by a box containing 13358 CHARMM-modified TIP3P \cite{tip32,tip3mod} water molecules. 
(The equilibrated system is a cube of edge length $\approx 73.5$~{\AA}. The starting equilibrated configuration was kindly provided by 
Karandur and Pettitt \cite{Karandur2015a}, who had simulated the system for over 100~ns at a temperature of  
300 K and a pressure of 1 atm.\ using, respectively, a Langevin thermostat and a Langevin barostat \cite{feller:jcp95}.) We maintained
the simulation parameters as in the Karandur-Pettitt study. Specifically, the barostat piston period was 100~fs and the decay time was 50~fs. 
The decay constant of the thermostat was 4 ps$^{-1}$. The SHAKE algorithm was used to constrain the geometry of water molecules and fix the bond 
between hydrogens and parent heavy atoms. Lennard-Jones interactions were terminated at 12.00 {\AA} by smoothly switching
to zero starting at 10.0 {\AA}. Electrostatic interactions were treated with the particle mesh Ewald method with a grid spacing of 1.0 {\AA}.
In contrast to the Karandur-Pettitt study, here we use a 2.0~fs timestep. \textit{In vacuo} calculations for peptide provided the vacuum reference. 
These \emph{in vacuo} simulations lasted at least 25 ns with a 1~fs timestep. The decay constant of the thermostat was 10 ps$^{-1}$.

To calculate the potential of mean force (PMF), $W(r)$, where  the order parameter $r$ is the distance between the terminal carbon atoms of the Gly$_{15}$ peptide, 
we first obtained one frame each with $r \in (30,40)$~{\AA} (domain L40), $r \in (25,35)$~{\AA} (domain L35), and $r \in (20,30)$~{\AA} (domain L30)
from the earlier simulations by Karandur and Pettitt \cite{Karandur2015a}. Then the PMFs in the respective domains were obtained using the adaptive-bias force (ABF) technique \cite{abf1,abf2}. 
Briefly, in the ABF approach, the order parameter  is binned in windows of width 0.1~{\AA} and using these counts initial biasing forces are estimated 
that encourage a uniform sampling of the order parameter in the chosen domain. As the simulation progresses, the distribution of $r$ and hence also
the biasing forces are updated. At convergence, the biasing force should cancel the force due to the underlying free energy surface (the quantity of interest), 
thus allowing the calculation of $W(r)$. 

For each domain,  ABF simulations spanned 26~ns. The first 16~ns was set aside for equilibration, during which time we monitored the evolution of the
biasing forces. Then the gradient of $W(r)$ obtained at the end of 18, 20, 22, 24, and 26~ns was averaged. The forces from the overlapping 
segments in L30 and L35 were averaged. The L30-L35 average and forces from L40 were then averaged to construct the gradient of $W(r)$ in the 
entire domain $r \in [20,40]$~{\AA}. The gradient
was then numerically integrated (a trapezoidal rule suffices) to obtained $W(r)$ from $r = 20.1$~{\AA} to $r = 39.9$~{\AA}. (For the \textit{in vacuo} ABF simulation,
we follow a similar procedure with gradients obtained at the end of 10, 15, 20, and 25~ns.) The potential energy of the peptide (in the solvent)
as a function of $r$ was obtained from the last 4~ns of the ABF trajectory and sorted and binned in windows of width 0.1~{\AA} along $r$. 
For the potential energy calculation, we used structures only from L40 and L30 simulations.

The calculation of $\mu^{\rm ex}$ and its entropic $Ts^{\rm ex}$ and enthalpic $h^{\rm ex}$ follows earlier work \cite{tomar:bj2013,tomar:jpcb14,tomar:jpcb16}. For completeness, the calculation approach is briefly described in the Appendix.

\section{Results}
\subsection{Free energy of chain compaction}
\begin{figure*}[h!]
\includegraphics[width=3.125in]{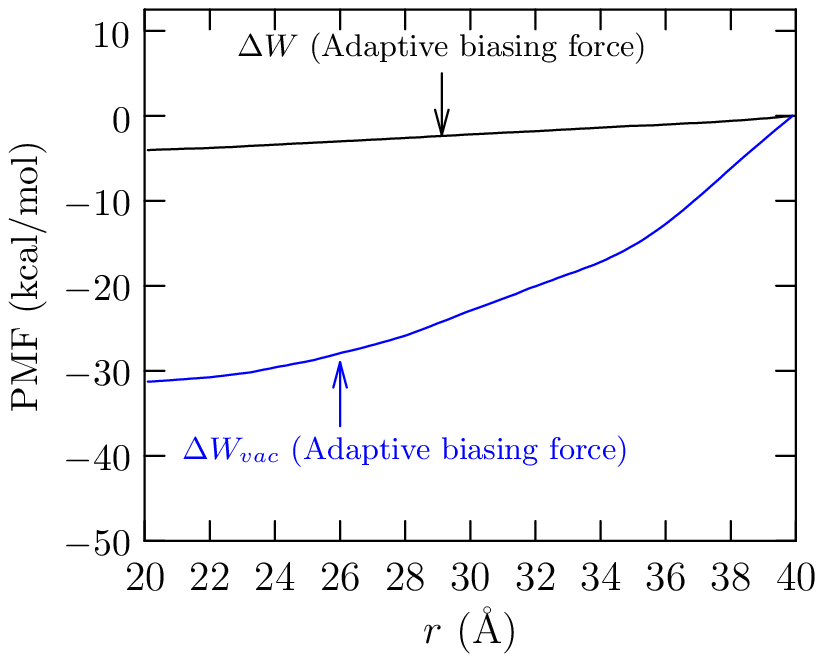}\hspace{3mm}\includegraphics[width=3.125in]{pmf_components}
\caption{\underline{Left panel}: PMF, $\Delta W(r)$, relative to $r = 40$~{\AA} for Gly$_{15}$ folding in the domain $r \in [20,40]$~{\AA}. PMF obtained from ABF is shown as solid black line. 
 \underline{Right panel}: The change in the internal energy $\Delta E$ (red line) and its electrostatic (blue) and van~der~Waals (grey) contributions for the solvated peptide. The light red shading indicates the $1\sigma$ standard error of the mean. The contributions from dihedral and angle
terms of the forcefield are negligible on the scale of the graph.}
\label{fg:pmf}
\end{figure*}

Figure~\ref{fg:pmf} shows the potential of mean force (PMF) between the terminal carbon atoms of the Gly$_{15}$. 
As $r$ decreases from 40~{\AA} to 20~{\AA}, the radius
of gyration of the peptide changes from about 13~{\AA} to about 6~{\AA}, indicating that the polypeptide adopts a compact configuration
as $r$ decreases. Figure~\ref{fg:pmf} shows that chain compaction is favored by a free energy change of approximately $-4$~kcal/mol.  Observe that there is an intrinsic drive for the peptide chain to collapse, as is seen in the potential of mean force for chain compaction
obtained in the absence of the solvent ($\Delta W_{vac}$) and as can also be inferred from the large intra-peptide energy change accompanying chain compaction
(Fig.~\ref{fg:pmf}, right panel).

\subsection{Analysis of intra-peptide interactions}

Given their role in organized structures such as the $\alpha$-helix and the $b$-sheet, it is natural to suspect that hydrogen bonds would contribute to the 
favorable intra-peptide electrostatic interaction, as has been suggested in earlier experimental studies  \cite{Kiefhaber:pnas06,Teufel:2011jmb}. 
It is standard practice, for example see Ref.\ \citenum{Tran:2008bk}, to identify hydrogen bonds on the basis of a geometric criterion. However, to obtain a
better understanding of the role of hydrogen bonds in the electrostatic contribution, which is of first interest here, it is also necessary to evaluate their energetic contribution. To this end, we analyzed hydrogen bonding contributions using both geometric and energetic criteria. 

\begin{figure}
\includegraphics[width=3.125in]{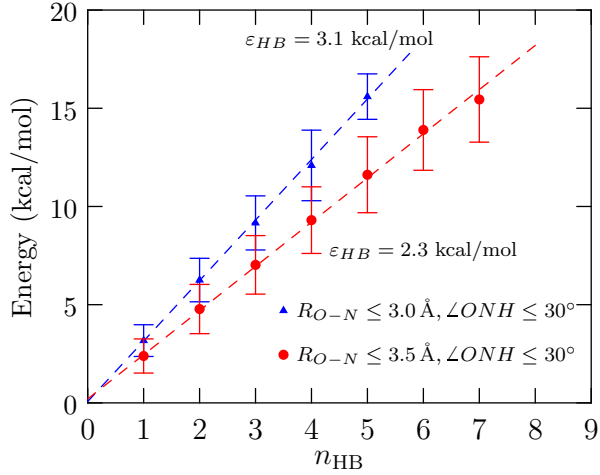}
\caption{Analysis of hydrogen bonding strength for hydrogen bonds defined according to various geometric criteria. $\varepsilon_{HB}$ is the
slope of the line and defines the average contribution of the single hydrogen bond based on the specified geometric criterion. The role of the end-caps, which can
also hydrogen bond, is ignored. Standard error of the mean is shown at $1\sigma$.}\label{fg:hbanalysis}
\end{figure}

First for all the sampled configurations (in $r \in [20,40]$~{\AA}), we calculated the number of hydrogen bonds based on the distance $r_c$ between the carbonyl 
oxygen for residue $i$ and the amide nitrogen (N) at $j$ and the angle $\theta_c$ between the N-H (amide proton) vector and the N-O vector. (For assessing
hydrogen bonds, $i$ and $j$ differ by at least 2 residues.) For a hydrogen bond pair satisfying the defined cutoffs, we find the pair interaction energy between the [CO]$_i$ group and the $[{\rm HNC_\alpha (H_\alpha)_2}]_j$ group. (Our choice of interacting groups is based on the fact that within the CHARMM forcefield, the CO group is neutral as is the $[{\rm HNC_\alpha (H_\alpha)_2}]$ group, but the bare NH group is not.) Figure~\ref{fg:hbanalysis} collects the results of this analysis for two commonly used cutoffs.  
 
 Figure~\ref{fg:hbanalysis} shows that the net interaction energy energy is linear in the number of hydrogen bond for both the defined cutoffs
 (Fig.~\ref{fg:hbanalysis}). Thus we can conclude that for the given forcefield, a hydrogen bond based on $r_c \leq 3.0$~{\AA} and
 $\theta_c \leq 30^\circ$ contributes \emph{on average} 3.1~kcal/mol favorably to the net binding strength. Likewise, a hydrogen bond based on 
 $r_c \leq 3.5$~{\AA} and $\theta_c \leq 30^\circ$ contributes \emph{on average} 2.3~kcal/mol to the net binding strength; it should be clear, however, 
 that this average includes the effect of the stronger hydrogen bonds that occur at the $3.1$~kcal/mol energy scale.

\begin{figure*}[h!]
\includegraphics[width=3.125in]{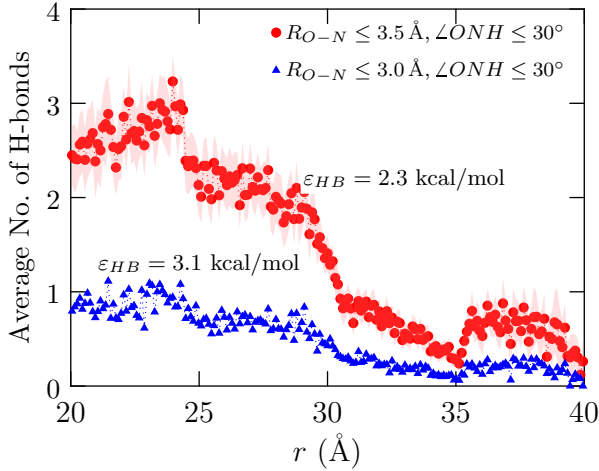}
\caption{Average number of hydrogen bonds as a function of the order parameter $r$ for various distance and angle criteria. (The role of the end-caps, which can
also hydrogen bond, is ignored.)  For clarity, 
standard error at $2\sigma$ is shown as a shaded background only for $R_{O-N} \leq 3.5\, {\rm {\AA}}, \angle ONH \leq 30^\circ$.}
\label{fg:hbonds}
\end{figure*}

Figure~\ref{fg:hbonds} shows that the average number of hydrogen bonds increases as $r$ (and hence also $R_g$, Fig.~\ref{fg:hbondsRg}) decreases, as has also been suggested experimentally \cite{Kiefhaber:pnas06,Teufel:2011jmb}.  
On average about 3 hydrogen bonds form upon collapse for the criterion $R_{ON} \leq 3.5$~{\AA} and $\angle ONH \leq 30^\circ$. Based on the analysis in Fig.~\ref{fg:hbanalysis}, we can infer than one of these is a H-bond contributing about 3.1~kcal/mol to the binding energy and the remaining two contribute
about 1.9~kcal/mol (on average) to the binding energy. 

\begin{figure*}[h!]
\includegraphics[width=3.125in]{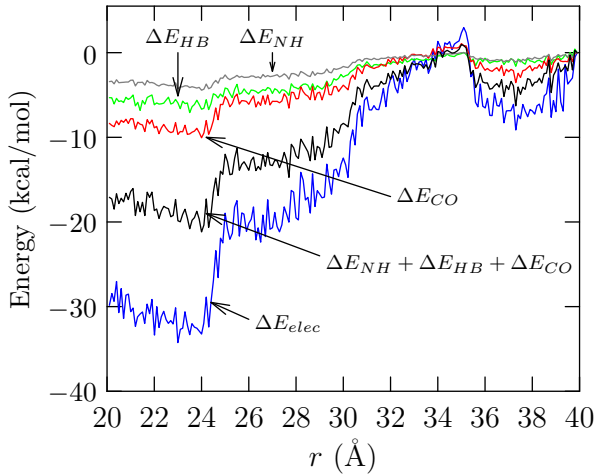}
\caption{Analysis of contributions to the electrostatic contribution to the internal energy change. Only energetic contribution from hydrogen bonds satisfying
$r_c \leq 3.5$~{\AA} and $\theta_c\leq 30^\circ$ is shown. The contribution from CO$_i$-CO$_j$ ($j \geq i + 2$) pairs
and NH$_i$-NH$_j$ ($j \geq i + 2$)  pairs are also indicated. For interactions involving NH, we always consider the neutral 
$[{\rm HNC_\alpha (H_\alpha)_2}]$ group. The role of the end-caps in H-bonding, CO-CO, and NH-NH interactions is ignored.  }\label{fg:elec}
\end{figure*}

Figure~\ref{fg:elec} compares the contribution from hydrogen bonds as well as due to interaction between CO-CO dipoles
and NH-NH dipoles. 
The results reveal that correlations between CO-groups play a  larger role in the net electrostatic energy change than hydrogen bonds
(based on $r_c \leq 3.5$~{\AA} and $\theta_c\leq 30^\circ$, $\varepsilon_{HB} = 2.3$~kcal/mol). Our identification of the importance of CO-CO
interactions is consistent with what has been reported earlier by Karandur~et~al.\  \cite{Karandur2014,Karandur2015a}.
However, in variance with their conclusion, we find that $\Delta E_{HB}$ also makes a significant contribution to the net electrostatic energy 
change. In particular, we find that $\Delta E_{HB}$ is about 63\% of $\Delta E_{CO}$. In a similar vein, we find that correlations between NH groups also
contributes favorably to the change in electrostatic energy. The sum of CO-CO, H-bonding, and NH-NH interactions is about 66\% 
of the net electrostatic change. For simplicity we have not included the interactions involving the terminal caps, which can participate
in all the three categories noted in Fig.~\ref{fg:elec}. Further, in comparing with the molecular dynamics data (Fig.~\ref{fg:elec}), 
we have ignored short range interaction involving partial charges that are not readily classifiable into one of the three defined categories 
noted in Fig.~\ref{fg:elec}. These contributions that have been left out contribute the rest of the change in $\Delta E_{elec}$. 

Summarizing the results of our analysis on intra-peptide interactions, we find that correlations between CO-groups and hydrogen bonds are two of the most important
contributions to the favorable change in $\Delta E_{elec}$. The identified importance of hydrogen bonds is also in good agreement with expectations based
on experiments \cite{Kiefhaber:pnas06,Teufel:2011jmb}.

\subsection{Role of hydration}

We next consider the analysis of hydration effects. To parse the effect of hydration, we write 
\begin{eqnarray}
\Delta W = \Delta W_{vac} + \Delta W_{ss} \, ,
\label{eq:dw}
\end{eqnarray}
where $\Delta W_{ss}$ accounts for all the hydration effects.  
Here $\Delta W_{ss} = \mu^{\rm ex} (r = 20.1) - \mu^{\rm ex} (r = 39.9)$, where $\mu^{\rm ex} (r)$ is the hydration free energy of the polypeptide with the constraint that the
end-to-end distance is $r$.  To estimate $\mu^{\rm ex} (r)$, we first classify the ensemble of conformations satisfying the constraint $r$ by the radius of gyration $R_g$.
For a given $r$, denoting the excess chemical potential of a specific conformation $R_g$  by $\mu^{\rm ex}( R_g | r)$, the multistate generalization \cite{lrp:cpms,lrp:book,merchant:jcp09,dixitpd:bj09} of the chemical potential $\mu^{\rm ex} (r)$ gives 
\begin{eqnarray}
\beta \mu^{\rm ex}(r) = \ln  \int_{R_g} e^{\beta \mu^{\rm ex} (R_g | r)} P(R_g | r) d R_g \, ,
\label{eq:muex}
\end{eqnarray}
where the integration is over all the conformations (classified according to $R_g$) that satisfy the constraint of fixed $r$, and $\beta = 1/k_{\rm B}T$, with $k_{\rm B}$ the Boltzmann constant and $T$ the temperature. $P(R_g|r) dR_g$ is probability of finding a conformation in the range $[R_g,R_g+dR_g]$ given the constraint $r$. 

\begin{figure}[h!]
\includegraphics{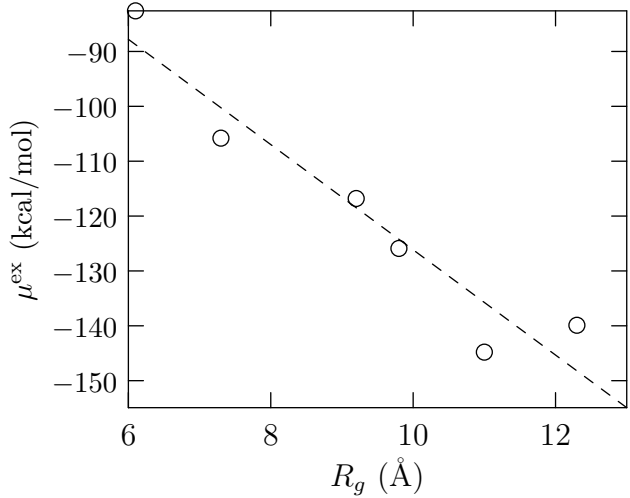}
\caption{Hydration free energy values for several Gly$_{15}$ conformations for $R_g$ values of interest in the present study. The linear fit is solely to
indicate that on average $\mu^{\rm ex}$ decreases with increasing $R_g$, i.e.\ as more of the chain is exposed to the solvent.} 
\label{fg:muexrg}
\end{figure}

Constructing $\mu^{\rm ex}(r)$ by calculating $\mu^{\rm ex} (R_g | r)$ for an ensemble of configurations is a daunting task, but much progress can be made using Eq.~\ref{eq:muex} and some physically realistic assumptions.  First we note that hydration free energy calculations for several different conformations of Gly$_{15}$ shows that $\mu^{\rm ex}$ for a given conformation is negative (Fig.~\ref{fg:muexrg}). 
This negative $\mu^{\rm ex}$ is also consistent with explicit hydration free energy calculations on shorter polyglycines \cite{tomar:bj2013,tomar:jpcb14} 
and is as expected based on hydration free energy calculations of another homogeneous peptides of varying chain lengths (up to about 10), for example, 
see \citenum{helms:2005fw,Hu:proteins2010,pettitt:jacs11,Kokubo:jpcb13,Harris:pnas14,tomar:jpcb16}.

Since $\mu^{\rm ex} (R_g | r) < 0$, it is clear that $\mu^{\rm ex}(r)$ must be bounded from above  by the least negative and from below by the most negative hydration free energy. Further since $\mu^{\rm ex} (R_g | r)$ decreases with increasing $R_g$, i.e.\ with increasing solvent exposure of the backbone, 
we can infer that for a given $r$, the hydration free energy $\mu^{\rm ex} (R_g | r)$ for the most collapsed conformation is expected to be least negative. 
Denoting the most collapsed conformation by $R^*_g$, we thus expect $[\mu^{\rm ex} (R_g | r) - \mu^{\rm ex}(R^*_g|r)] \leq 0$ and thus 
\begin{eqnarray}
\beta \mu^{\rm ex}(r) & = &   \beta \mu^{\rm ex} (R^*_g | r) +  \ln  \int_{R_g} e^{\beta [\mu^{\rm ex} (R_g | r) - \mu^{\rm ex}(R^*_g|r)]} P(R_g | r) d R_g \nonumber \\
& \leq & \beta \mu^{\rm ex} (R^*_g | r) \, .
\label{eq:muex1}
\end{eqnarray}

For using Eq.~\ref{eq:muex1}, we first obtained two structures satisfying $r=39.9$~{\AA} and $r=20.1$~{\AA}, respectively, from the ABF trajectory. (We find 
a structure that is within 0.05~{\AA} of the target distance and then adjust $r$.) Subsequently, these peptide configurations were centered and rotated such that the
end-to-end vector is along the principal diagonal of the simulation cell. With the terminal carbon atoms fixed in space, we sampled conformations of the
peptide from 2~ns of production. 

\begin{figure}
\includegraphics{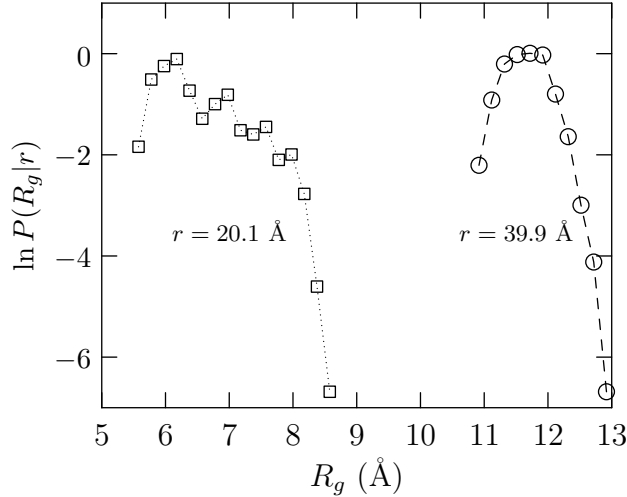}
\caption{Probability distribution of $R_g$ values for the specified end-to-end distance. For $r = 20.1$~{\AA}, the $R_g$ of the most collapsed
conformation is 5.5~{\AA} and for $r = 39.9$~{\AA}, the $R_g$ of the most collapsed conformation is 10.8~{\AA}. These $R_g$ values
fall slightly to the left of the leftmost point shown in the plot.}
\label{fg:prg}
\end{figure}

Analysis of the distribution of $R_g$ for $r=20.1$~{\AA} and $39.9$~{\AA}, shows that $P(R_g^* | r) \approx e^{-2}$ relative to the most probable $\bar{R}_g$, i.e.
$-\ln [P(R^*_g | r) / P(\bar{R}_g | r)] \approx 2\, {\rm k_{\rm B}T}$ (Fig.~\ref{fg:prg}).
But for the same increase in $R_g$, about 1~{\AA}, the hydration free energy decreases by $O(17\, {\rm k_{\rm B}T})$ (Fig.~\ref{fg:muexrg}). 
Because of the exponential dependence of the free energy on $[\mu^{\rm ex} (R_g | r) - \mu^{\rm ex}(R^*_g|r)] < 0$ which decreases sharply
relative to the growth in $P(R_g|r)$, we expect the upper bound to itself be a fair approximation to the required free energy. (See also
Ref.~\citenum{dixitpd:bj09}  for a similar argument in the context of ion hydration.) Thus, we expect that 
the hydration contribution in Eq.~\ref{eq:dw} can be approximated as 
\begin{eqnarray}
\Delta W_{ss} & = & \mu^{\rm ex}[r = 20.1~{\rm {\AA}}] - \mu^{\rm ex}[r = 39.9~{\rm {\AA}}] \nonumber \\
 & \approx &  \mu^{\rm ex}(R^*_g | r = 20.1~{\rm {\AA}}) - \mu^{\rm ex} (R^*_g | r = 39.9~{\rm {\AA}})
 \label{eq:approx}
 \end{eqnarray}

For the $(R^*_g | r = 20.1)$ and $(R^*_g | r= 39.9)$ structures, we find the hydration free energy, $\mu^{\rm ex}$, using the regularization approach to hydration free energies \cite{weber:jcp11,Weber:jctc12,tomar:bj2013,tomar:jpcb14} (Appendix A), a technique that is based on the extensively documented quasichemical organization of the potential distribution theorem \cite{lrp:book,lrp:cpms}.  As before \cite{tomar:jpcb14,tomar:jpcb16}, we also obtained the entropic ($s^{\rm ex}$) and enthalpic ($h^{\rm ex}$) decomposition of $\mu^{\rm ex}$ (Appendix A). 

Table~\ref{tb:muexdata} collects the results of the hydration analysis and it is clear that the calculated value of the free energy of collapse is in reasonable
accord with the value obtained using the ABF procedure (Fig.~\ref{fg:pmf}).  
\begin{table}
\caption{Hydration and intra-peptide interaction contributions in the collapse of Gly$_{15}$ from $r = 39.9$~{\AA} to $r = 20.1$~{\AA}. $\Delta W_{ss}$ is based
on Eq.~\ref{eq:approx}. $\Delta W ({\rm calc.}) = \Delta W_{ss} + W_{vac}$ is the value of the free energy of collapse using the calculated hydration free energy;
$\Delta W$ (ABF) is the corresponding value from Fig.~\ref{fg:pmf}.}\label{tb:muexdata}
\begin{tabular}{l | r} 
 Quantity & (kcal/mol)  \\ \hline
 $\Delta W_{vac}$ (ABF) & $-31.3$ \\
 $\Delta W_{ss}$ & $25.5 \pm 2$ \\
 $\Delta W$ (calc.) & $-5.8 \pm 2$ \\
 $\Delta W$ (ABF) & $-4.0$ \\ \hline
 \end{tabular}
 \end{table}

\begin{figure*}[h!]
\includegraphics[scale=0.95]{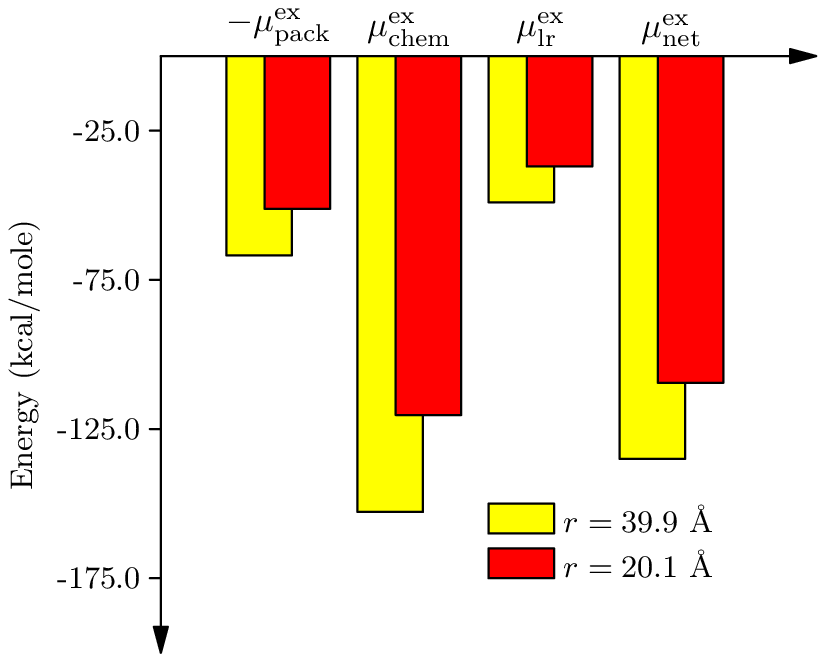}\hspace{2mm}\includegraphics[scale=0.95]{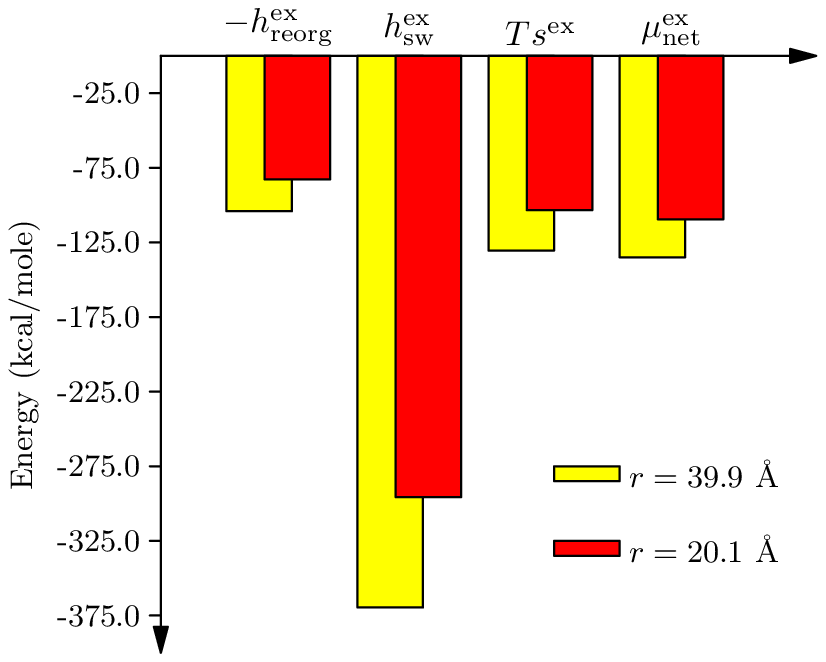}
\caption{\underline{Left panel}: The hydration free energy and its components based on the quasichemical decompostion \cite{lrp:book,lrp:cpms,weber:jcp11,Weber:jctc12,tomar:bj2013,tomar:jpcb14}. (See Appendix.)  The packing contribution measures the free energy to create a cavity to accommodate the peptide \cite{tomar:jpcb14,tomar:jpcb16}; it is the domain that is excluded to the solvent centers and is uniquely defined for the given forcefield.  The chemistry contribution captures the role of solute attractive interactions with solvent in the first hydration shell, here defined to be the surface at a distance of 5~{\AA} away from nearest heavy atom in the solute. The long-range contribution is the free energy of interaction between the peptide and the solvent when solvent is excluded from the first hydration layer.  \underline{Right panel}: Decomposition of the hydration free energy change into enthalpic and entropic contributions. The enthalpy of hydration is further separated into a solvent reorganization and solute-solvent interaction parts.}
\label{fg:muexqc}
\end{figure*}

Analyzing $\Delta W_{ss}$ shows that the packing contribution, a measure of primitive hydrophobic effects\cite{Pratt:1992p3019,Pratt:2002p3001}, does favor chain compaction, as is expected (Fig.~\ref{fg:muexqc}).  
But this packing contribution is approximately balanced by the long-range contributions that 
favor chain unfolding.  Importantly, the chemistry contribution reflecting the role of favorable solute interactions with the solvent in the first hydration shell
is nearly twice the magnitude of the packing contribution and favors chain unfolding. \emph{Thus hydrophilic effects overwhelm hydrophobic effects to shift the balance to the unfolded state.} 

Mirroring the packing contribution, the energetic cost to reorganize the solvent around a cavity ($h^{\rm ex}_{reorg}$)  favors chain compaction, as does the 
entropy of hydration. But favorable solute-water interactions reflected in $h^{\rm ex}_{sw}$ greatly favor chain expansion. This observation suggests that the backbone must play a substantial role in protein folding, consistent with several recent studies \cite{Rose:pnasbackbone,Bolen:rev08,Auton:bc11}.

\section{Conclusions}

We find that the collapse of Gly$_{15}$ is driven by intra-molecular interactions,  which  are primarily electrostatic in origin. The basis for this electrostatic
drive is found mostly in favorable CO-CO interactions, hydrogen bonding interactions between CO and NH 
groups, and also interaction between amide group (NH) dipoles. Favorable solute-solvent interaction dominates the hydration thermodynamics and 
opposes the collapse of Gly$_{15}$, despite packing (or primitive hydrophobic) effects favoring chain compaction. The net balance between intramolecular interactions and hydration is such that intramolecular contributions win by a small margin and drive the collapse of the peptide. Thus liquid water is both a good solvent for the hydration of the peptide unit \cite{Hu:proteins2010,tomar:bj2013}, but also a poor solvent from the perspective of folding, as
the hydration effects lose in comparison to intra-peptide interactions. Our work suggests that the hydration of the peptide backbone is likely an important 
determinant in the solution thermodynamics of intrinsically disordered peptides, an aspect that needs to be investigated further. 
The observed feature of hydration opposing collapse driven by favorable intramolecular interactions is also expected to be relevant to protein folding and assembly.

\begin{acknowledgements}
This research used resources of the National Energy Research Scientific Computing Center, which is supported by the Office of Science of the U.S. Department of Energy under Contract No. DE- AC02-05CH11231. 

\end{acknowledgements}

\newpage
\section{Appendix}
\setcounter{section}{0}
\makeatletter
\renewcommand{\thesection}{S.\@Roman\c@section}

\setcounter{figure}{0}
\makeatletter
\renewcommand{\thefigure}{S\@arabic\c@figure}

\setcounter{table}{0}
\makeatletter
\renewcommand{\thetable}{S.\@Roman\c@table}

\setcounter{equation}{0}
\makeatletter
\renewcommand{\theequation}{S.\@arabic\c@equation}

\section{Methods}
The free energy of hydration, $\mu^{\rm ex}$, is given as 
\begin{eqnarray}
\beta \mu^{\rm ex} = \underbrace{\ln x_0(\lambda)}_{\rm chemistry} \underbrace{- \ln p_0(\lambda)}_{\rm packing} + \underbrace{\beta\mu^{\rm ex} (n = 0 | \lambda)}_{\rm long-range}
\label{eq:qc}
\end{eqnarray}
within the quasichemical organization of the potential distribution theorem \cite{lrp:book,lrp:cpms}. Each of the terms in the above equation has a simple
physical interpretation, as has been noted before \cite{tomar:bj2013,tomar:jpcb14}. 

In Eq.~\ref{eq:qc}, $\lambda$ is the distance to which solvent is excluded from the surface of the solute in computing the chemical contribution to hydration. Typically, excluding
the solvent in the first hydration shell ($\lambda \approx 5$~{\AA}) suffices. This choice also ensures that the binding energy distribution of the solute with the 
solvent outside the defined inner-shell is Gaussian to a good approximation (see below). 

The largest value of $\lambda$, labelled $\lambda_{\rm SE}$, for which the chemistry contribution is zero has a special meaning. It demarcates the domain 
within which solvent cannot enter, i.e.\ the solvent is excluded.  For the given forcefield, this surface is uniquely defined.  We find that $\lambda_{\rm SE} \approx 3$~{\AA}. With this choice, Eq.~\ref{eq:qc} can be rearranged
as, 
\begin{eqnarray}
\beta \mu^{\rm ex} & = &  \underbrace{ \ln \left[x_0(\lambda)\frac{p_0 (\lambda_{\rm SE})}{p_0(\lambda)}\right]}_{\rm renormalized\, chemistry} \underbrace{-\ln p_0 (\lambda_{\rm SE})}_{\rm SE\, packing} +  \underbrace{\beta\mu^{\rm ex} (n = 0 | \lambda)}_{\rm long-range}
\label{eq:qc1}
\end{eqnarray}
The term identified as renormalized chemistry has the following physical meaning. It is the work done to move the solvent interface a distance $\lambda$ away from the solute relative to the case when the only role played by the solute is to exclude solvent up to $\lambda_{\rm SE}$.  This term illuminates the role of short-range solute-solvent attractive interactions on hydration. This decomposition is different from the ones we have used in the past  \cite{tomar:bj2013,tomar:jpcb14}. The results in the present study are based  on Eq.~\ref{eq:qc1}.

\subsection{Chemistry and packing contributions}
We apply atom-centered fields to carve out a molecular cavity in the liquid \cite{tomar:bj2013,tomar:jpcb14,tomar:jpcb16}.  We use the Tcl-interface to NAMD \cite{namd} to impose forces on the solvent due to the field.  The functional form of the field was 
as before (Eq.~4b, Ref.~\citenum{Weber:jctc12}):
\begin{eqnarray}
\phi_\lambda (r) & = & 4a\bigg[ \left(\frac{b}{r-\lambda+\sqrt[6]{2}b}\right)^{12} - \left(\frac{b}{r-\lambda+\sqrt[6]{2}b}\right)^{6} \bigg] \nonumber \\
& + & a \label{eq:lj} \; ,
\end{eqnarray}
where $a = 0.155$~kcal/mol and $b = 3.1655$~{\AA} are positive constants and  ($r < \lambda$), and $\phi_\lambda (r) = 0$ for $r \geq \lambda$.

To build the field to its eventual range of $\lambda = 5$~{\AA}, we progressively apply the field, and for every unit {\AA} increment in the range, 
we compute the work done in applying the field using a seven-point Gauss-Legendre quadrature \cite{Hummer:jcp96}. In earlier studies we have used a 5-point quadrature. The 
calculated values using 5- and 7-points are the same within statistical uncertainties, but a 7-point quadrature allows us to use fewer number of time steps
per point (here 0.9~ns versus 1.2~ns in earlier studies).   The following seven Gauss-points \\ $\left[0, \pm 0.4058,  \pm 0.7415, \pm 0.9491\right]$ are chosen for each unit {\AA}. At each Gauss-point, the system was simulated for 0.9~ns and the (force) data from the last 0.5~ns used for analysis. (Excluding more data did not change the numerical value significantly, indicating good convergence.) Error analysis and error propagation was performed as before \cite{Weber:jctc12}: the standard error of the mean force was obtained using the Friedberg-Cameron algorithm \cite{friedberg:1970,allen:error} and in adding multiple quantities, the errors were propagated using standard variance-addition rules. 

The starting configuration for each $\lambda$ point is obtained from the ending configuration of the previous point in the chain of states. For the packing contributions, a total of 35 Gauss points span $\lambda \in [0,5]$.  For the chemistry contribution, since solvent never enters $\lambda < 2.5$~{\AA}, we simulate $\lambda \in [2,5]$ for a total of 21 Gauss points. 

\subsection{Long-range contribution}

Let the conditional solute-solvent binding energy distribution be $P(\varepsilon|\phi_\lambda)$ and the solute-solvent binding energy distribution with solute and solvent thermally uncoupled be $P^{(0)}(\varepsilon|\phi_\lambda)$. For a large enough conditioning radius, we expect both these distributions to be well described by a gaussian.  
Then \cite{lrp:book,lrp:cpms}
\begin{eqnarray}
\mu^{\rm ex} [P(\varepsilon|\phi_\lambda)] & = & \langle \varepsilon |\phi_\lambda \rangle + \frac{\beta}{2}\sigma^2 \nonumber \\
\mu^{\rm ex} [P^{(0)}(\varepsilon|\phi_\lambda)] & = & \langle \varepsilon | \phi_\lambda \rangle_0 - \frac{\beta}{2}\sigma^2 \, .
\label{eq:gaussian}
\end{eqnarray}
In the above equations, $\langle \varepsilon |\phi_\lambda \rangle$ and $ \langle \varepsilon |\phi_\lambda \rangle_0$ are the mean binding energies in the coupled and uncoupled ensembles, respectively, and $\sigma^2$ is the variance of the distribution, the same for both $P(\varepsilon|\phi_\lambda)$ and $P^{(0)}(\varepsilon|\phi_\lambda)$. 

For characterizing  $P(\varepsilon|\phi_\lambda)$ (with $\lambda = 5$~{\AA}), the starting configuration for the $\lambda = 5$~{\AA} simulation was obtained from the endpoint of the Gauss-Legendre procedure for the chemistry calculation; for $P^{(0)}(\varepsilon|\phi_\lambda)$ (with $\lambda = 5$~{\AA}), we use the neat solvent state at the endpoint of the packing 
calculation. The system was equilibrated for 0.9~ns and data collected over an additional 1.2~ns with configurations saved every 0.5~ps.  Protein solvent binding energies
were obtained using the {\sc PairInteraction} module in NAMD.

\begin{figure}[h!]
\includegraphics{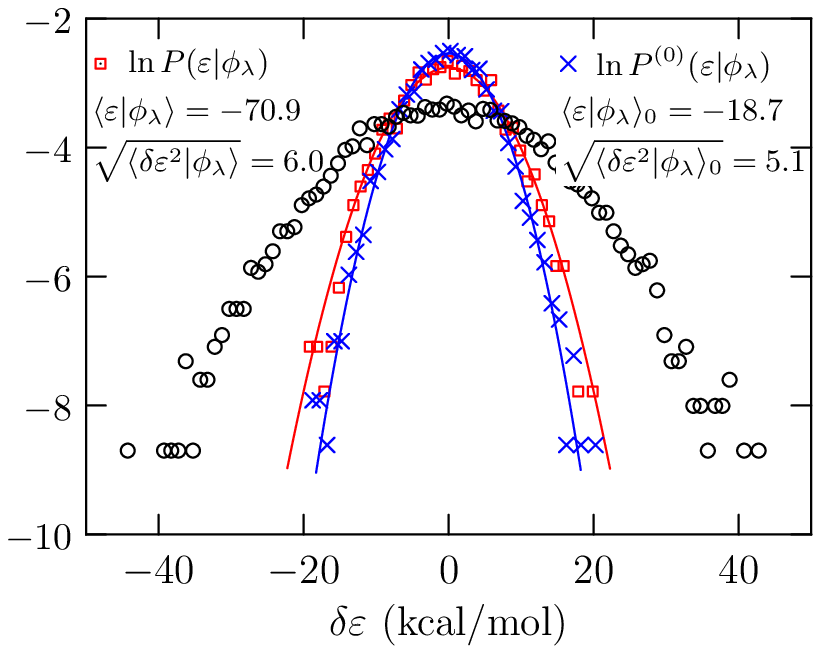}
\caption{The coupled and uncoupled regularized  binding energy distributions for the case where the separation between the terminal carbon atoms is 20.1~{\AA} and
the structure has the smallest $R_g$. The coupled (but not regularized) distribution is shown using open circles ($\circ$). Regularization serves to reduce the
variance of this distribution. Not that despite the seemingly gaussian behavior, the high-energy tail region of the non-regularized distribution is not well characterized; it is in fact expected to obey an extreme value distribution \cite{asthagiri:jcp2008}.}
\label{fg:bedist}
\end{figure}

Figure~\ref{fg:bedist} shows that as expected the $P(\varepsilon|\phi_\lambda)$ and $P^{(0)}(\varepsilon|\phi_\lambda)$ distributions are gaussian. For this particular system,
however, the variance is slightly different for these distributions. [The origins of this behavior lie in the fact that the partial charges of the peptide backbone are largely unshielded from the solvent. For example, were the backbone to be decorated with apolar groups, as happens for a polyalanine, the conditioned coupled and uncoupled distribution have the same variance \cite{tomar:jpcb16}.]
Nevertheless, $\mu^{\rm ex} [P(\varepsilon|\phi_\lambda)] = -40.5\pm 1.0$~kcal/mol is in excellent agreement with $\mu^{\rm ex} [P^{(0)}(\varepsilon|\phi_\lambda)] = -40.4\pm 1.0$~kcal/mole. These numbers are also in excellent agreement with $-41.8$~kcal/mol obtained using the regularization approach for vdW interactions and a 2-point
Gauss-Legendre quadrature for electrostatics \cite{Weber:jctc12}. The estimate $\mu^{\rm ex}(n=0|\lambda) = 0.5\cdot (\langle \varepsilon |\phi_\lambda \rangle +  \langle \varepsilon | \phi_\lambda \rangle_0) = -44.8\pm 0.2$  (cf.\ Eq.~\ref{eq:gaussian}; see also Ref.~\citenum{beck:jcp08})), which should be valid if both distributions are strictly gaussian,
is also in error from the quadrature result by only about 7\%.   

Similar analysis was also performed for the extended state of the peptide. The final results reported are based on $\mu^{\rm ex} [P(\varepsilon|\phi_\lambda)]$, which is particularly easy to obtain from the conditioned-peptide simulations.

\section{Enthalpic and entropic contributions to hydration}

From the Euler relation for the pure solvent and the solvent with one added solute, we can show that the excess entropy of hydration
is  
\begin{eqnarray}
T  s^{\rm ex} &  = & E^{\rm ex} - kT^2 \alpha_p + p(\langle V^{\rm ex} \rangle + kT \kappa_T) - \mu^{\rm ex} \nonumber \\ 
	              & \approx & E_{sw} + E_{reorg}  - \mu^{\rm ex}
\label{eq:entropy}	              
\end{eqnarray}
where $\kappa_T$ is the isothermal compressibility and $\alpha_p$ is the thermal expansivity of the solvent. The average excess energy of hydration, $E^{\rm ex}$,  is the sum the average solute-water interaction energy $E_{sw}$ and $E_{reorg}$, the reorganization energy. The latter is given by the change in the average potential energy of the solvent in the solute-solvent system minus that in the neat solvent system. (Note that solute-solvent interactions  are not counted as part of $E_{reorg}$.)  Ignoring pressure-volume effects, the excess enthalpy of hydration $h^{\rm ex} =  E^{\rm ex}$. The solute-solvent interaction contribution $E_{sw}$ can be further decomposed into backbone-solvent, $E_{bb}$, and sidechain-solvent, $E_{sc}$, contributions. These contributions were straightforwardly obtained using the {\sc PairInteraction} module within NAMD. The coupled peptide solvent system was simulated for an additional 3~ns and frames were archived every 500~fs for interaction-energy analysis. 

For calculating $E_{reorg}$ we adapted the hydration-shell-wise procedure developed earlier \cite{asthagiri:jcp2008}. We 
define an inner-shell around the peptide as the union of shells of radius $\lambda$ centered  on the peptide heavy atoms.  $\lambda \leq 5.5$~{\AA}, $5.5 < \lambda \leq 8.5$~{\AA}, and $8.5 < \lambda \leq 11.5$~{\AA}  defined the first, second, and third shells, respectively. For the reorganization calculation, the definition of the inner shell was slightly increased by 0.5~{\AA}, but this change has no bearing on the final thermodynamic quantity $h^{\rm ex}$. 
Let $n_w$ be the number of water molecules in a shell for some chosen configuration.  The potential energy of these $n_w$ waters is given by the interaction energy between these $n_w$ waters plus half the interaction energy of these $n_w$ waters with the rest of the fluid. We thus find the average potential energy, $\langle E_{shell}\rangle $, and the average population, $\langle n_{shell}\rangle $, for a given shell.  The contribution to the average reorganization energy from the shell is then  $\langle E_{shell}\rangle - \langle n_{shell}\rangle \cdot \langle \varepsilon_w\rangle $. Errors are propagated using standard rules. 

For all cases, we find that by the third shell bulk behavior is attained; that is, $E_{reorg, 3} \approx 0$ within statistical uncertainties, where $E_{reorg, 3}$ is the reorganization energy contribution from the third (3$^{rd}$) shell. 

\section{Distribution of hydrogen bonds versus $R_g$}

\begin{figure*}[h!]
\includegraphics{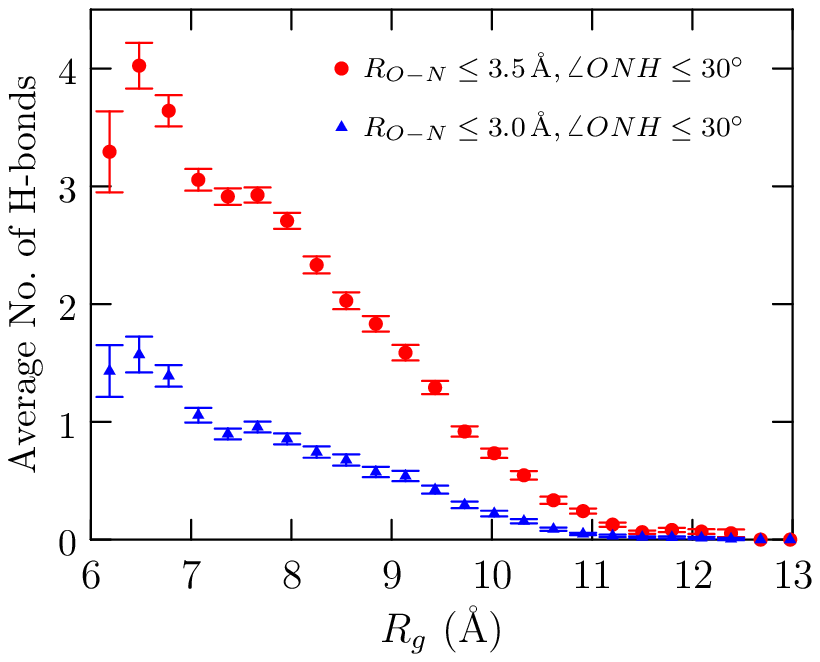}
\caption{Average number of hydrogen bonds as a function of the radius of gyration $R_g$ and for various distance and angle criteria. The $R_g$ range is divided into 25 bins and configurations sorted into the appropriate bins. The mean number of hydrogen bonds from configurations in a bin and the associated standard error is then obtained using standard relations.}
\label{fg:hbondsRg}
\end{figure*}
The slight difference in the average counts versus $R_g$  and average counts versus $r$ (Fig.~\ref{fg:hbonds}) occurs because the relation between $r$  and $R_g$ is itself subject to some statistical uncertainty. Thus sorting configurations using $r$ or  $R_g$ as order parameters can influence the averaging of the dependent variable (here the number of hydrogen bonds). However, the physical conclusion that number of hydrogen bonds increases upon chain collapse is independent 
of these considerations.

\newpage


\begin{thebibliography}{42}
\expandafter\ifx\csname natexlab\endcsname\relax\def\natexlab#1{#1}\fi
\expandafter\ifx\csname bibnamefont\endcsname\relax
  \def\bibnamefont#1{#1}\fi
\expandafter\ifx\csname bibfnamefont\endcsname\relax
  \def\bibfnamefont#1{#1}\fi
\expandafter\ifx\csname citenamefont\endcsname\relax
  \def\citenamefont#1{#1}\fi
\expandafter\ifx\csname url\endcsname\relax
  \def\url#1{\texttt{#1}}\fi
\expandafter\ifx\csname urlprefix\endcsname\relax\def\urlprefix{URL }\fi
\providecommand{\bibinfo}[2]{#2}
\providecommand{\eprint}[2][]{\url{#2}}

\bibitem[{\citenamefont{Kauzmann}(1959)}]{Kauzmann:59}
\bibinfo{author}{\bibfnamefont{W.}~\bibnamefont{Kauzmann}},
  \bibinfo{journal}{Adv. Prot. Chem.} \textbf{\bibinfo{volume}{14}},
  \bibinfo{pages}{1} (\bibinfo{year}{1959}).

\bibitem[{\citenamefont{Chandler}(2005)}]{chandler:nature05}
\bibinfo{author}{\bibfnamefont{D.}~\bibnamefont{Chandler}},
  \bibinfo{journal}{Nature} \textbf{\bibinfo{volume}{437}},
  \bibinfo{pages}{640} (\bibinfo{year}{2005}).

\bibitem[{\citenamefont{Dill}(1990)}]{dill:1990ww}
\bibinfo{author}{\bibfnamefont{K.~A.} \bibnamefont{Dill}},
  \bibinfo{journal}{Biochem.} \textbf{\bibinfo{volume}{29}},
  \bibinfo{pages}{7133} (\bibinfo{year}{1990}).

\bibitem[{\citenamefont{Dill and MacCallum}(2012)}]{Dill2012}
\bibinfo{author}{\bibfnamefont{K.~A.} \bibnamefont{Dill}} \bibnamefont{and}
  \bibinfo{author}{\bibfnamefont{J.~L.} \bibnamefont{MacCallum}},
  \bibinfo{journal}{Science} \textbf{\bibinfo{volume}{338}},
  \bibinfo{pages}{1042} (\bibinfo{year}{2012}).

\bibitem[{\citenamefont{Teufel et~al.}(2011)\citenamefont{Teufel, Johnson, Lum,
  and Neuweiler}}]{Teufel:2011jmb}
\bibinfo{author}{\bibfnamefont{D.~P.} \bibnamefont{Teufel}},
  \bibinfo{author}{\bibfnamefont{C.~M.} \bibnamefont{Johnson}},
  \bibinfo{author}{\bibfnamefont{J.~K.} \bibnamefont{Lum}}, \bibnamefont{and}
  \bibinfo{author}{\bibfnamefont{H.}~\bibnamefont{Neuweiler}},
  \bibinfo{journal}{J. Mol. Biol.} \textbf{\bibinfo{volume}{409}},
  \bibinfo{pages}{250} (\bibinfo{year}{2011}).

\bibitem[{\citenamefont{Tran et~al.}(2008)\citenamefont{Tran, Mao, and
  Pappu}}]{Tran:2008bk}
\bibinfo{author}{\bibfnamefont{H.~T.} \bibnamefont{Tran}},
  \bibinfo{author}{\bibfnamefont{A.}~\bibnamefont{Mao}}, \bibnamefont{and}
  \bibinfo{author}{\bibfnamefont{R.~V.} \bibnamefont{Pappu}},
  \bibinfo{journal}{J. Am. Chem. Soc.} \textbf{\bibinfo{volume}{130}},
  \bibinfo{pages}{7380} (\bibinfo{year}{2008}).

\bibitem[{\citenamefont{Hu et~al.}(2010{\natexlab{a}})\citenamefont{Hu, Lynch,
  Kokubo, and Pettitt}}]{Hu:2010b}
\bibinfo{author}{\bibfnamefont{C.~Y.} \bibnamefont{Hu}},
  \bibinfo{author}{\bibfnamefont{G.~C.} \bibnamefont{Lynch}},
  \bibinfo{author}{\bibfnamefont{H.}~\bibnamefont{Kokubo}}, \bibnamefont{and}
  \bibinfo{author}{\bibfnamefont{B.~M.} \bibnamefont{Pettitt}},
  \bibinfo{journal}{Proteins: Struc. Func. Bioinform.}
  \textbf{\bibinfo{volume}{78}}, \bibinfo{pages}{695}
  (\bibinfo{year}{2010}{\natexlab{a}}).

\bibitem[{\citenamefont{Cornette et~al.}(1987)\citenamefont{Cornette, Cease,
  Margalit, Spouge, Berzofsky, and DeLisi}}]{delisi:jmb87}
\bibinfo{author}{\bibfnamefont{J.~L.} \bibnamefont{Cornette}},
  \bibinfo{author}{\bibfnamefont{K.~B.} \bibnamefont{Cease}},
  \bibinfo{author}{\bibfnamefont{H.}~\bibnamefont{Margalit}},
  \bibinfo{author}{\bibfnamefont{J.~L.} \bibnamefont{Spouge}},
  \bibinfo{author}{\bibfnamefont{J.~A.} \bibnamefont{Berzofsky}},
  \bibnamefont{and} \bibinfo{author}{\bibfnamefont{C.}~\bibnamefont{DeLisi}},
  \bibinfo{journal}{J. Mol. Biol.} \textbf{\bibinfo{volume}{195}},
  \bibinfo{pages}{659} (\bibinfo{year}{1987}).

\bibitem[{\citenamefont{Wilce et~al.}(1995)\citenamefont{Wilce, Aguilar, and
  Hearn}}]{Wilce1995}
\bibinfo{author}{\bibfnamefont{M.~C.~J.} \bibnamefont{Wilce}},
  \bibinfo{author}{\bibfnamefont{M.-I.} \bibnamefont{Aguilar}},
  \bibnamefont{and} \bibinfo{author}{\bibfnamefont{M.~T.~W.}
  \bibnamefont{Hearn}}, \bibinfo{journal}{Anal.Chem.}
  \textbf{\bibinfo{volume}{67}}, \bibinfo{pages}{1210} (\bibinfo{year}{1995}).

\bibitem[{\citenamefont{M{\"o}glich et~al.}(2006)\citenamefont{M{\"o}glich,
  Joder, and Kiefhaber}}]{Kiefhaber:pnas06}
\bibinfo{author}{\bibfnamefont{A.}~\bibnamefont{M{\"o}glich}},
  \bibinfo{author}{\bibfnamefont{K.}~\bibnamefont{Joder}}, \bibnamefont{and}
  \bibinfo{author}{\bibfnamefont{T.}~\bibnamefont{Kiefhaber}},
  \bibinfo{journal}{Proc. Natl. Acad. Sc. USA} \textbf{\bibinfo{volume}{103}},
  \bibinfo{pages}{12394} (\bibinfo{year}{2006}).

\bibitem[{\citenamefont{Karandur et~al.}(2014)\citenamefont{Karandur, Wong, and
  Pettitt}}]{Karandur2014}
\bibinfo{author}{\bibfnamefont{D.}~\bibnamefont{Karandur}},
  \bibinfo{author}{\bibfnamefont{K.-Y.} \bibnamefont{Wong}}, \bibnamefont{and}
  \bibinfo{author}{\bibfnamefont{B.~M.} \bibnamefont{Pettitt}},
  \bibinfo{journal}{J. Phys. Chem. B} \textbf{\bibinfo{volume}{118}},
  \bibinfo{pages}{9565} (\bibinfo{year}{2014}).

\bibitem[{\citenamefont{Karandur et~al.}(2015)\citenamefont{Karandur, Harris,
  and Pettitt}}]{Karandur2015a}
\bibinfo{author}{\bibfnamefont{D.}~\bibnamefont{Karandur}},
  \bibinfo{author}{\bibfnamefont{R.~C.} \bibnamefont{Harris}},
  \bibnamefont{and} \bibinfo{author}{\bibfnamefont{B.~M.}
  \bibnamefont{Pettitt}}, \bibinfo{journal}{Prot. Sci.} pp.
  \bibinfo{pages}{103--110} (\bibinfo{year}{2015}).

\bibitem[{\citenamefont{Weber et~al.}(2011)\citenamefont{Weber, Merchant, and
  Asthagiri}}]{weber:jcp11}
\bibinfo{author}{\bibfnamefont{V.}~\bibnamefont{Weber}},
  \bibinfo{author}{\bibfnamefont{S.}~\bibnamefont{Merchant}}, \bibnamefont{and}
  \bibinfo{author}{\bibfnamefont{D.}~\bibnamefont{Asthagiri}},
  \bibinfo{journal}{J. Chem. Phys.} \textbf{\bibinfo{volume}{135}},
  \bibinfo{pages}{181101} (\bibinfo{year}{2011}).

\bibitem[{\citenamefont{Weber and Asthagiri}(2012)}]{Weber:jctc12}
\bibinfo{author}{\bibfnamefont{V.}~\bibnamefont{Weber}} \bibnamefont{and}
  \bibinfo{author}{\bibfnamefont{D.}~\bibnamefont{Asthagiri}},
  \bibinfo{journal}{J. Chem. Theory Comput.} \textbf{\bibinfo{volume}{8}},
  \bibinfo{pages}{3409} (\bibinfo{year}{2012}).

\bibitem[{\citenamefont{Tomar et~al.}(2014)\citenamefont{Tomar, Weber, Pettitt,
  and Asthagiri}}]{tomar:jpcb14}
\bibinfo{author}{\bibfnamefont{D.~S.} \bibnamefont{Tomar}},
  \bibinfo{author}{\bibfnamefont{V.}~\bibnamefont{Weber}},
  \bibinfo{author}{\bibfnamefont{B.~M.} \bibnamefont{Pettitt}},
  \bibnamefont{and}
  \bibinfo{author}{\bibfnamefont{D.}~\bibnamefont{Asthagiri}},
  \bibinfo{journal}{J. Phys. Chem. B} \textbf{\bibinfo{volume}{118}},
  \bibinfo{pages}{4080} (\bibinfo{year}{2014}).

\bibitem[{\citenamefont{Tomar et~al.}(2016)\citenamefont{Tomar, Weber, Pettitt,
  and Asthagiri}}]{tomar:jpcb16}
\bibinfo{author}{\bibfnamefont{D.~S.} \bibnamefont{Tomar}},
  \bibinfo{author}{\bibfnamefont{W.}~\bibnamefont{Weber}},
  \bibinfo{author}{\bibfnamefont{M.~B.} \bibnamefont{Pettitt}},
  \bibnamefont{and}
  \bibinfo{author}{\bibfnamefont{D.}~\bibnamefont{Asthagiri}},
  \bibinfo{journal}{J. Phys. Chem. B} \textbf{\bibinfo{volume}{120}},
  \bibinfo{pages}{69} (\bibinfo{year}{2016}).

\bibitem[{\citenamefont{Jorgensen et~al.}(1983)\citenamefont{Jorgensen,
  Chandrasekhar, $\ldots$, and Klein}}]{tip32}
\bibinfo{author}{\bibfnamefont{W.}~\bibnamefont{Jorgensen}},
  \bibinfo{author}{\bibfnamefont{J.}~\bibnamefont{Chandrasekhar}},
  \bibinfo{author}{\bibnamefont{$\ldots$}}, \bibnamefont{and}
  \bibinfo{author}{\bibfnamefont{M.~L.} \bibnamefont{Klein}},
  \bibinfo{journal}{J. Chem. Phys.} \textbf{\bibinfo{volume}{79}},
  \bibinfo{pages}{926} (\bibinfo{year}{1983}).

\bibitem[{\citenamefont{Neria et~al.}(1996)\citenamefont{Neria, Fischer, and
  Karplus}}]{tip3mod}
\bibinfo{author}{\bibfnamefont{E.}~\bibnamefont{Neria}},
  \bibinfo{author}{\bibfnamefont{S.}~\bibnamefont{Fischer}}, \bibnamefont{and}
  \bibinfo{author}{\bibfnamefont{M.}~\bibnamefont{Karplus}},
  \bibinfo{journal}{J. Chem. Phys.} \textbf{\bibinfo{volume}{105}},
  \bibinfo{pages}{1902} (\bibinfo{year}{1996}).

\bibitem[{\citenamefont{Feller et~al.}(1995)\citenamefont{Feller, Zhang,
  Pastor, and Brooks}}]{feller:jcp95}
\bibinfo{author}{\bibfnamefont{S.~E.} \bibnamefont{Feller}},
  \bibinfo{author}{\bibfnamefont{Y.}~\bibnamefont{Zhang}},
  \bibinfo{author}{\bibfnamefont{R.~W.} \bibnamefont{Pastor}},
  \bibnamefont{and} \bibinfo{author}{\bibfnamefont{B.~R.}
  \bibnamefont{Brooks}}, \bibinfo{journal}{J. Chem. Phys.}
  \textbf{\bibinfo{volume}{103}}, \bibinfo{pages}{4613} (\bibinfo{year}{1995}).

\bibitem[{\citenamefont{Darve et~al.}(2008)\citenamefont{Darve,
  Rodriguez-G{\'o}mez, and Pohorille}}]{abf1}
\bibinfo{author}{\bibfnamefont{E.}~\bibnamefont{Darve}},
  \bibinfo{author}{\bibfnamefont{D.}~\bibnamefont{Rodriguez-G{\'o}mez}},
  \bibnamefont{and}
  \bibinfo{author}{\bibfnamefont{A.}~\bibnamefont{Pohorille}},
  \bibinfo{journal}{J. Chem. Phys.} \textbf{\bibinfo{volume}{128}},
  \bibinfo{pages}{144120} (\bibinfo{year}{2008}).

\bibitem[{\citenamefont{H{\'e}nin et~al.}(2010)\citenamefont{H{\'e}nin, Fiorin,
  Chipot, and Klein}}]{abf2}
\bibinfo{author}{\bibfnamefont{J.}~\bibnamefont{H{\'e}nin}},
  \bibinfo{author}{\bibfnamefont{G.}~\bibnamefont{Fiorin}},
  \bibinfo{author}{\bibfnamefont{C.}~\bibnamefont{Chipot}}, \bibnamefont{and}
  \bibinfo{author}{\bibfnamefont{M.~L.} \bibnamefont{Klein}},
  \bibinfo{journal}{J. Chem. Theory Comput.} \textbf{\bibinfo{volume}{6}},
  \bibinfo{pages}{35} (\bibinfo{year}{2010}).

\bibitem[{\citenamefont{Tomar et~al.}(2013)\citenamefont{Tomar, Weber, and
  Asthagiri}}]{tomar:bj2013}
\bibinfo{author}{\bibfnamefont{D.~S.} \bibnamefont{Tomar}},
  \bibinfo{author}{\bibfnamefont{V.}~\bibnamefont{Weber}}, \bibnamefont{and}
  \bibinfo{author}{\bibfnamefont{D.}~\bibnamefont{Asthagiri}},
  \bibinfo{journal}{Biophys. J.} \textbf{\bibinfo{volume}{105}},
  \bibinfo{pages}{1482} (\bibinfo{year}{2013}).

\bibitem[{\citenamefont{Pratt and Asthagiri}(2007)}]{lrp:cpms}
\bibinfo{author}{\bibfnamefont{L.~R.} \bibnamefont{Pratt}} \bibnamefont{and}
  \bibinfo{author}{\bibfnamefont{D.}~\bibnamefont{Asthagiri}}, in
  \emph{\bibinfo{booktitle}{Free energy calculations: {Theory} and applications
  in chemistry and biology}}, edited by
  \bibinfo{editor}{\bibfnamefont{C.}~\bibnamefont{Chipot}} \bibnamefont{and}
  \bibinfo{editor}{\bibfnamefont{A.}~\bibnamefont{Pohorille}}
  (\bibinfo{publisher}{Springer}, \bibinfo{address}{Berlin, DE},
  \bibinfo{year}{2007}), vol.~\bibinfo{volume}{86} of
  \emph{\bibinfo{series}{Springer series in {Chemical Physics}}},
  chap.~\bibinfo{chapter}{9}, pp. \bibinfo{pages}{323--351}.

\bibitem[{\citenamefont{Beck et~al.}(2006)\citenamefont{Beck, Paulaitis, and
  Pratt}}]{lrp:book}
\bibinfo{author}{\bibfnamefont{T.~L.} \bibnamefont{Beck}},
  \bibinfo{author}{\bibfnamefont{M.~E.} \bibnamefont{Paulaitis}},
  \bibnamefont{and} \bibinfo{author}{\bibfnamefont{L.~R.} \bibnamefont{Pratt}},
  \emph{\bibinfo{title}{The potential distribution theorem and models of
  molecular solutions}} (\bibinfo{publisher}{Cambridge University Press},
  \bibinfo{address}{Cambridge, UK}, \bibinfo{year}{2006}).

\bibitem[{\citenamefont{Merchant and Asthagiri}(2009)}]{merchant:jcp09}
\bibinfo{author}{\bibfnamefont{S.}~\bibnamefont{Merchant}} \bibnamefont{and}
  \bibinfo{author}{\bibfnamefont{D.}~\bibnamefont{Asthagiri}},
  \bibinfo{journal}{J. Chem. Phys.} \textbf{\bibinfo{volume}{130}},
  \bibinfo{pages}{195102} (\bibinfo{year}{2009}).

\bibitem[{\citenamefont{Dixit et~al.}(2009)\citenamefont{Dixit, Merchant, and
  Asthagiri}}]{dixitpd:bj09}
\bibinfo{author}{\bibfnamefont{P.~D.} \bibnamefont{Dixit}},
  \bibinfo{author}{\bibfnamefont{S.}~\bibnamefont{Merchant}}, \bibnamefont{and}
  \bibinfo{author}{\bibfnamefont{D.}~\bibnamefont{Asthagiri}},
  \bibinfo{journal}{Biophys. J.} \textbf{\bibinfo{volume}{96}},
  \bibinfo{pages}{2138} (\bibinfo{year}{2009}).

\bibitem[{\citenamefont{Staritzbichler
  et~al.}(2005)\citenamefont{Staritzbichler, Gu, and Helms}}]{helms:2005fw}
\bibinfo{author}{\bibfnamefont{R.}~\bibnamefont{Staritzbichler}},
  \bibinfo{author}{\bibfnamefont{W.}~\bibnamefont{Gu}}, \bibnamefont{and}
  \bibinfo{author}{\bibfnamefont{V.}~\bibnamefont{Helms}}, \bibinfo{journal}{J.
  Phys. Chem. B} \textbf{\bibinfo{volume}{109}}, \bibinfo{pages}{19000}
  (\bibinfo{year}{2005}).

\bibitem[{\citenamefont{Hu et~al.}(2010{\natexlab{b}})\citenamefont{Hu, Kokubo,
  Lynch, Bolen, and Pettitt}}]{Hu:proteins2010}
\bibinfo{author}{\bibfnamefont{C.~Y.} \bibnamefont{Hu}},
  \bibinfo{author}{\bibfnamefont{H.}~\bibnamefont{Kokubo}},
  \bibinfo{author}{\bibfnamefont{G.}~\bibnamefont{Lynch}},
  \bibinfo{author}{\bibfnamefont{D.~W.} \bibnamefont{Bolen}}, \bibnamefont{and}
  \bibinfo{author}{\bibfnamefont{B.~M.} \bibnamefont{Pettitt}},
  \bibinfo{journal}{Prot. Sc.} \textbf{\bibinfo{volume}{19}},
  \bibinfo{pages}{1011} (\bibinfo{year}{2010}{\natexlab{b}}).

\bibitem[{\citenamefont{Kokubo et~al.}(2011)\citenamefont{Kokubo, Hu, and
  Pettitt}}]{pettitt:jacs11}
\bibinfo{author}{\bibfnamefont{H.}~\bibnamefont{Kokubo}},
  \bibinfo{author}{\bibfnamefont{C.~Y.} \bibnamefont{Hu}}, \bibnamefont{and}
  \bibinfo{author}{\bibfnamefont{B.~M.} \bibnamefont{Pettitt}},
  \bibinfo{journal}{J. Am. Chem. Soc.} \textbf{\bibinfo{volume}{133}},
  \bibinfo{pages}{1849} (\bibinfo{year}{2011}).

\bibitem[{\citenamefont{Kokubo et~al.}(2013)\citenamefont{Kokubo, Harris,
  Asthagiri, and Pettitt}}]{Kokubo:jpcb13}
\bibinfo{author}{\bibfnamefont{H.}~\bibnamefont{Kokubo}},
  \bibinfo{author}{\bibfnamefont{R.~C.} \bibnamefont{Harris}},
  \bibinfo{author}{\bibfnamefont{D.}~\bibnamefont{Asthagiri}},
  \bibnamefont{and} \bibinfo{author}{\bibfnamefont{B.~M.}
  \bibnamefont{Pettitt}}, \bibinfo{journal}{J. Phys. Chem. B}
  \textbf{\bibinfo{volume}{117}}, \bibinfo{pages}{16428}
  (\bibinfo{year}{2013}).

\bibitem[{\citenamefont{Harris and Pettitt}(2014)}]{Harris:pnas14}
\bibinfo{author}{\bibfnamefont{R.~C.} \bibnamefont{Harris}} \bibnamefont{and}
  \bibinfo{author}{\bibfnamefont{B.~M.} \bibnamefont{Pettitt}},
  \bibinfo{journal}{Proc. Natl. Acad. Sc. USA} \textbf{\bibinfo{volume}{111}},
  \bibinfo{pages}{14681} (\bibinfo{year}{2014}).

\bibitem[{\citenamefont{Pratt and Pohorille}(1992)}]{Pratt:1992p3019}
\bibinfo{author}{\bibfnamefont{L.~R.} \bibnamefont{Pratt}} \bibnamefont{and}
  \bibinfo{author}{\bibfnamefont{A.}~\bibnamefont{Pohorille}},
  \bibinfo{journal}{Proc. Natl. Acad. Sc. USA} \textbf{\bibinfo{volume}{89}},
  \bibinfo{pages}{2995} (\bibinfo{year}{1992}).

\bibitem[{\citenamefont{Pratt}(2002)}]{Pratt:2002p3001}
\bibinfo{author}{\bibfnamefont{L.~R.} \bibnamefont{Pratt}},
  \bibinfo{journal}{Ann. Rev. Phys. Chem.} \textbf{\bibinfo{volume}{53}},
  \bibinfo{pages}{409} (\bibinfo{year}{2002}).

\bibitem[{\citenamefont{Rose et~al.}(2006)\citenamefont{Rose, Fleming, Banavar,
  and Maritan}}]{Rose:pnasbackbone}
\bibinfo{author}{\bibfnamefont{G.~D.} \bibnamefont{Rose}},
  \bibinfo{author}{\bibfnamefont{P.~J.} \bibnamefont{Fleming}},
  \bibinfo{author}{\bibfnamefont{J.~R.} \bibnamefont{Banavar}},
  \bibnamefont{and} \bibinfo{author}{\bibfnamefont{A.}~\bibnamefont{Maritan}},
  \bibinfo{journal}{Proc. Natl. Acad. Sc. USA} \textbf{\bibinfo{volume}{103}},
  \bibinfo{pages}{16623} (\bibinfo{year}{2006}).

\bibitem[{\citenamefont{Bolen and Rose}(2008)}]{Bolen:rev08}
\bibinfo{author}{\bibfnamefont{D.~W.} \bibnamefont{Bolen}} \bibnamefont{and}
  \bibinfo{author}{\bibfnamefont{G.~D.} \bibnamefont{Rose}},
  \bibinfo{journal}{Annu. Rev. Biochem.} \textbf{\bibinfo{volume}{77}},
  \bibinfo{pages}{339} (\bibinfo{year}{2008}).

\bibitem[{\citenamefont{Auton et~al.}(2011)\citenamefont{Auton, R{\"o}sgen,
  Sinev, Holthauzen, and Bolen}}]{Auton:bc11}
\bibinfo{author}{\bibfnamefont{M.}~\bibnamefont{Auton}},
  \bibinfo{author}{\bibfnamefont{J.}~\bibnamefont{R{\"o}sgen}},
  \bibinfo{author}{\bibfnamefont{M.}~\bibnamefont{Sinev}},
  \bibinfo{author}{\bibfnamefont{L.~M.} \bibnamefont{Holthauzen}},
  \bibnamefont{and} \bibinfo{author}{\bibfnamefont{D.~W.} \bibnamefont{Bolen}},
  \bibinfo{journal}{Biophys. Chem.} \textbf{\bibinfo{volume}{159}},
  \bibinfo{pages}{90} (\bibinfo{year}{2011}).

\bibitem[{\citenamefont{Kale et~al.}(1999)\citenamefont{Kale, Skeel,
  Bhandarkar, Brunner, Gursoy, Krawetz, Phillips, Shinozaki, Varadarajan, and
  Schulten}}]{namd}
\bibinfo{author}{\bibfnamefont{L.}~\bibnamefont{Kale}},
  \bibinfo{author}{\bibfnamefont{R.}~\bibnamefont{Skeel}},
  \bibinfo{author}{\bibfnamefont{M.}~\bibnamefont{Bhandarkar}},
  \bibinfo{author}{\bibfnamefont{R.}~\bibnamefont{Brunner}},
  \bibinfo{author}{\bibfnamefont{A.}~\bibnamefont{Gursoy}},
  \bibinfo{author}{\bibfnamefont{N.}~\bibnamefont{Krawetz}},
  \bibinfo{author}{\bibfnamefont{J.}~\bibnamefont{Phillips}},
  \bibinfo{author}{\bibfnamefont{A.}~\bibnamefont{Shinozaki}},
  \bibinfo{author}{\bibfnamefont{K.}~\bibnamefont{Varadarajan}},
  \bibnamefont{and} \bibinfo{author}{\bibfnamefont{K.}~\bibnamefont{Schulten}},
  \bibinfo{journal}{J. Comput. Phys.} \textbf{\bibinfo{volume}{151}},
  \bibinfo{pages}{283} (\bibinfo{year}{1999}).

\bibitem[{\citenamefont{Hummer and Szabo}(1996)}]{Hummer:jcp96}
\bibinfo{author}{\bibfnamefont{G.}~\bibnamefont{Hummer}} \bibnamefont{and}
  \bibinfo{author}{\bibfnamefont{A.}~\bibnamefont{Szabo}}, \bibinfo{journal}{J.
  Chem. Phys.} \textbf{\bibinfo{volume}{105}}, \bibinfo{pages}{2004}
  (\bibinfo{year}{1996}).

\bibitem[{\citenamefont{Friedberg and Cameron}(1970)}]{friedberg:1970}
\bibinfo{author}{\bibfnamefont{R.}~\bibnamefont{Friedberg}} \bibnamefont{and}
  \bibinfo{author}{\bibfnamefont{J.~E.} \bibnamefont{Cameron}},
  \bibinfo{journal}{J. Chem. Phys.} \textbf{\bibinfo{volume}{52}},
  \bibinfo{pages}{6049} (\bibinfo{year}{1970}).

\bibitem[{\citenamefont{Allen and Tildesley}(1987)}]{allen:error}
\bibinfo{author}{\bibfnamefont{M.~P.} \bibnamefont{Allen}} \bibnamefont{and}
  \bibinfo{author}{\bibfnamefont{D.~J.} \bibnamefont{Tildesley}},
  \emph{\bibinfo{title}{Computer simulation of liquids}}
  (\bibinfo{publisher}{Oxford University Press}, \bibinfo{year}{1987}), chap.
  \bibinfo{chapter}{6. How to analyze the results}, pp.
  \bibinfo{pages}{192--195}.

\bibitem[{\citenamefont{Asthagiri et~al.}(2008)\citenamefont{Asthagiri,
  Merchant, and Pratt}}]{asthagiri:jcp2008}
\bibinfo{author}{\bibfnamefont{D.}~\bibnamefont{Asthagiri}},
  \bibinfo{author}{\bibfnamefont{S.}~\bibnamefont{Merchant}}, \bibnamefont{and}
  \bibinfo{author}{\bibfnamefont{L.~R.} \bibnamefont{Pratt}},
  \bibinfo{journal}{J. Chem. Phys.} \textbf{\bibinfo{volume}{128}},
  \bibinfo{pages}{244512} (\bibinfo{year}{2008}).

\bibitem[{\citenamefont{Rogers and Beck}(2008)}]{beck:jcp08}
\bibinfo{author}{\bibfnamefont{D.~M.} \bibnamefont{Rogers}} \bibnamefont{and}
  \bibinfo{author}{\bibfnamefont{T.~L.} \bibnamefont{Beck}},
  \bibinfo{journal}{J. Chem. Phys.} \textbf{\bibinfo{volume}{129}},
  \bibinfo{pages}{134505} (\bibinfo{year}{2008}).

\end{thebibliography}

\end{document}